# Medial Injury/Dysfunction Induced Granulation Tissue Repair is the Pathogenesis of Atherosclerosis

Xinggang Wang[1,2], Aijun Sun[1], Junbo Ge[1,]*

[1]Department of Cardiology, Zhongshan Hospital, Fudan University. Shanghai Institute of Cardiovascular Diseases, Shanghai, China

[2]Institute of Biomedical Sciences, Fudan University, Shanghai, China

*Corresponding author.

## Abstract

Atherosclerosis, a chronic lesion of vascular wall, remains a leading cause of death and loss of life years. Classical hypotheses for atherosclerosis are long-standing mainly to explain atherogenesis. Unfortunately, these hypotheses may not explain the variation in the susceptibility to atherosclerosis. These issues are controversial over the past 150 years. Atherosclerosis from human coronary arteries was examined and triangle of media was found to be a true portraiture of cells injury in the media, and triangle of intima was a true portraiture of myofibroblast proliferation, extracellular matrix (ECM) secretion, collagen fiber formation and intimal thickening to repair media dysfunction. Myofibroblasts, ECM and lumen (intima)/vasa vasorum (VV) (adventitia) constitute granulation tissue repair. With granulation tissue hyperplasia, lots of collagen fibers (normal or denatured), foam cells and new capillaries formed. Thus, the following theory was postulated: Risk factors induce smooth muscle cells (SMCs) injury/loss, and fibrosis or structure destruction could be developed in the media, which lead to media dysfunction. Media dysfunction prompts disturbed mechanical properties of blood vessels, resulting in bigger pressure buildup in the intima and adventitia. Granulation tissues in the intima/adventitia develop to repair the injured media. Atherosclerosis, stiffening or aneurysm develops depending on media dysfunction severity and granulated tissue repair mode/degree. Nearly all characteristics of clinical atherosclerosis could be ideally interpreted with the theory. We believe that media dysfunction is a key initiator in the pathogenesis of atherosclerosis. It is expected that media dysfunction theory of atherosclerosis, should offer better understanding of the etiology for atherosclerosis.

**Introduction**

Atherosclerosis is the most predominant causative factor for cardiovascular and cerebrovascular diseases, serious threats to human health. Although there are many hypotheses trying to explain its occurrence, they still could not explain the characteristics of clinical atherosclerosis well. Therefore, the pathogenesis behind atherosclerosis remains unclear. Many questions have been puzzling for a long time: As a continuous pipeline in human, 1. Sugar, lipids, oxygen, inflammatory factors, etc. are basically the same after simultaneous transplantation of saphenous vein grafts (SVGs) and radial artery grafts (RAGs) in coronary artery bypass grafting (CABG) patients. The patency of RAGs is much better than that of SVGs(1). 2. Why medium and large arteries, especially muscular arteries, are prone to atherogenesis, whereas arterioles and in situ veins are rarely seen with atherogenesis? 3. Atherosclerosis is more pronounced at the outer walls of bifurcations/convex surface of curved arteries compared with their neighborhoods. 4. Why myocardial bridges are less likely to be conflicted with atherosclerosis compared with non-myocardial bridges in a given individual? 5. Why "diabetic foot" is common, while "diabetic hand" is not common? Do they share a common pathogenesis? The long-standing hypotheses, e.g., intimal injury, lipid deposition, low shear stress, turbulence and inflammation, have been widely perceived to explain atherogenesis. Unfortunately, any given hypothesis cannot explain one or several phenomena. In fact, it was not well explained the variation in atherosclerosis susceptibility. Our perception to-date might fail to grasp the nature of atherosclerosis. We believe that there is only one truth and atherosclerosis should have a common pathogenesis instead of lots of hypotheses. Atherosclerosis from human coronary arteries was examined and compelling evidences had prompted a new theory raised here. All the characteristics of clinical atherosclerosis, which are confusing or unexplainable with the previous hypotheses, could be interpreted with our novel theory.

**Materials and methods**

    **Vascular tissues**

Human coronary arteries (n=66) were collected from autopsies of the Institute of Forensic Sciences of Shanghai, which conformed to the principles outlined in the Declaration of Helsinki and obtained written consent from their relatives prior to the inclusion of subjects in the study. The tissues were fixed with formalin, washed with water overnight, dehydrated with gradient alcohol, cleared by dimethylbenzene, embedded in paraffin, and sectioned into 4 μm thick on the slides.

**Hematoxylin-eosin (HE) staining**

The paraffin was removed from the section with dimethylbenzene. The slides were infiltrated with gradient alcohol, finally to water. The slices were stained in hematoxylin solution for 5 minutes, washed with water and ethanol hydrochloride, stained with eosin for 2 minutes. Then they were dehydrated with gradient alcohol, cleared with dimethylbenzene, and sealed with neutral gum.

**MOVAT staining**

The paraffin was removed from the section with dimethylbenzene. The slides were infiltrated with gradient alcohol, finally to water. The tissues were dyed with alcian blue for 15 minutes and washed with water for 2 minutes, differentiated with basic ethanol and washed with water, dyed with weigert hematoxylin and washed with water, dyed with Saffron o-acid fuchsin and washed with water, dyed with phosphomolybdic acid, dyed with glacial acetic acid and washed with water, dyed with sirius red and dehydrated with anhydrous alcohol, cleared with dimethylbenzene, and sealed with neutral gum.

**Whole slide imaging**

Whole slide imaging was performed with 3DHISTECH digital slice scanner. The images were captured with CaseViewer 2.3.

**Fluid physics**

This article involved many common knowledges of fluid physics(2). Bernoulli's equation and continuity equation of fluid($\rho Sv$=constant1) are also involved in this paper.

**Newton's Mechanics**

Newton's Mechanics, especially Newton's Third Law, is applied in the analysis of vascular wall mechanics.

**Results**

Smooth muscle cells (SMCs) injury/loss provoked fibrosis, muscular property loss and structure destruction of the medial layer.

Granulation tissue established in the intima/adventitia in an effort to repair the medial injury. Extracellular matrix (ECM) secretion, extracellular collagen fibers formation and angiogenesis were ubiquitous during granulation tissue repair.

There were lots of collagen fibers and foam cells in granulation tissue from the intimal layer. In addition, abundant denatured collagen fibers were distributed in the areas free of capillaries which is far from the lumen.

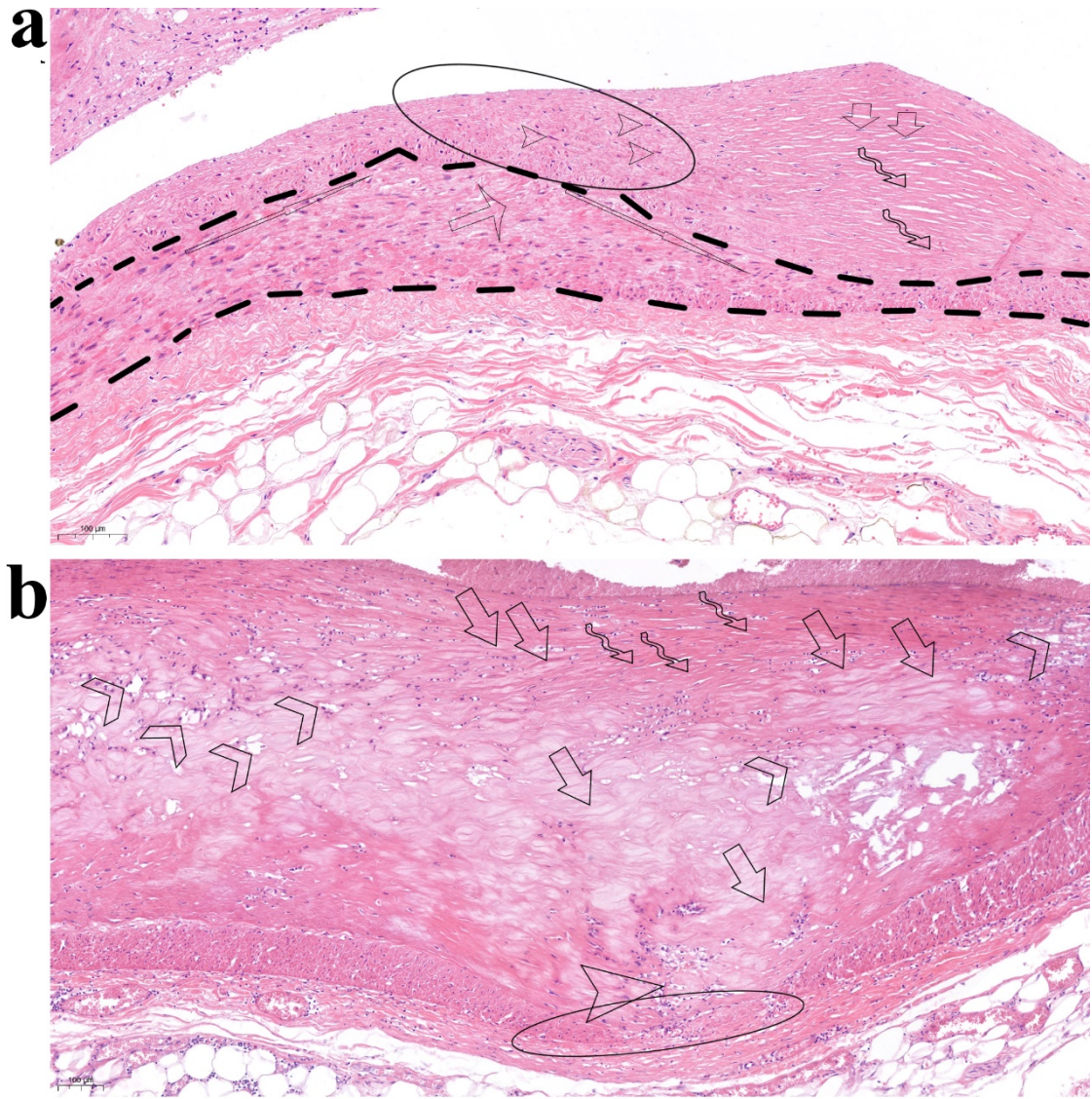

Fig.1 | Media dysfunction and granulation tissue repair inside

a: The triangle of media was a true portraiture of medial injury, while the triangle of intima was a true portraiture of myofibroblasts formation, secretion of extracellular matrix (ECM) and collagen fibers accumulation (fibrous plaque) in the intima which repaired the media. Myofibroblasts, ECM and the lumen constitute the granulation tissue in the intima (detail in Extended Data Fig. 1).

b: HE shows that thick granulation tissue repaired the serious destroyed media which lost structure and function. ECM secretion, extracellular collagen fibers formation and angiogenesis were ubiquitous in granulation tissue repair. There were lots of collagen fibers and foam cells in the fibers in granulation tissue. In addition, lots of denatured collagen fibers were distributed in the areas without capillaries and far from the lumen (soft plaque) (detail in Extended Data Fig. 2).

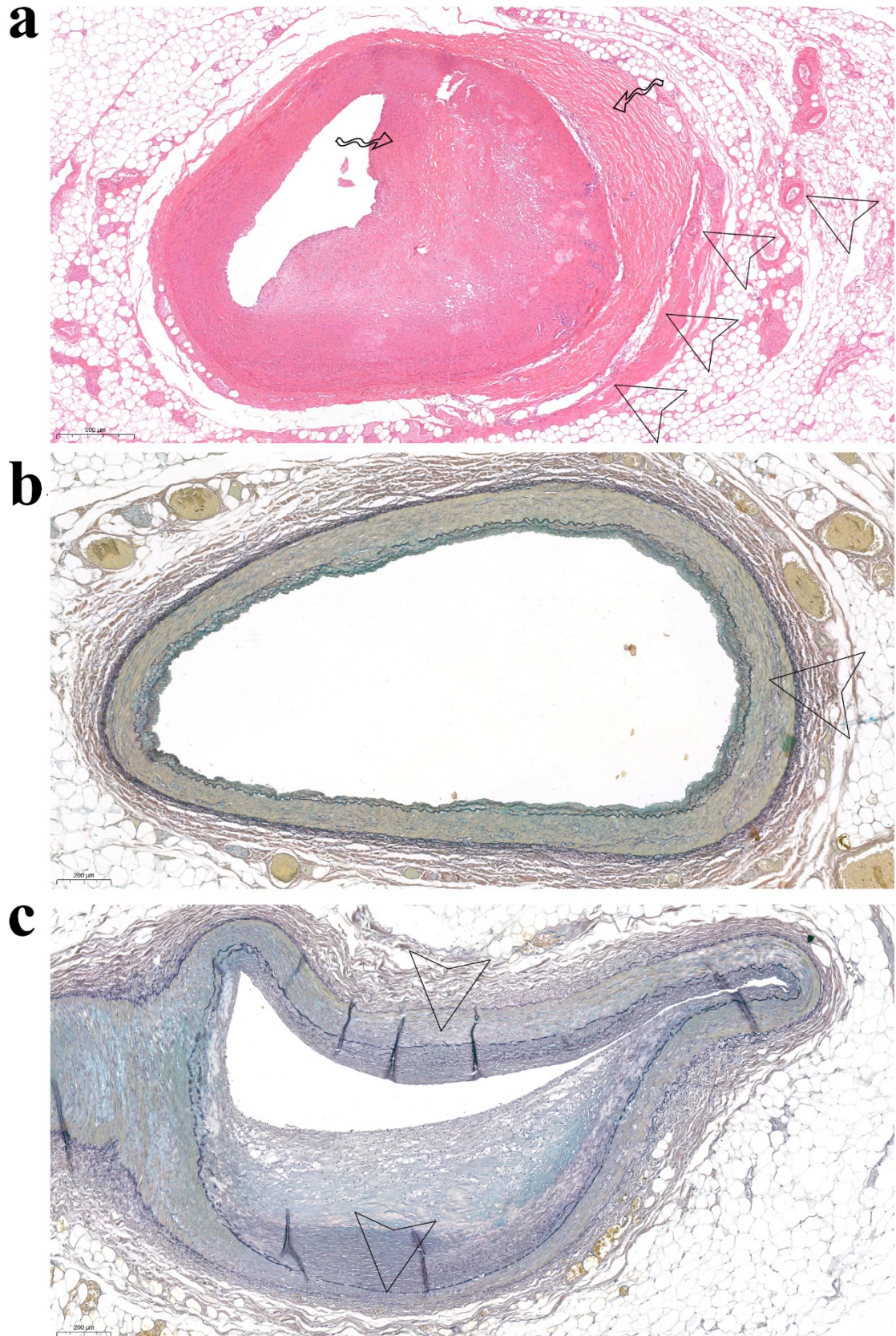

Fig.2 |

a: HE micrograph shows that the media was destroyed seriously and the lumen had severe stenosis. Granulation tissue repair could be seen in both intima and adventitia. There were many small vessels nourishing the granulation tissue in the adventitia (detail in Extended Data Fig. 3).

b: MOVAT staining showed that the media of normal coronary artery had good muscular property (detail in Extended Data Fig. 4a).

c: MOVAT staining showed that the media of coronary artery suffered from muscular property loss, fibrosis or structural damage. In the intima, there were fibrous plaques or soft plaques formed with the media dysfunction degree (detail in Extended Data Fig. 4b and Extended Data Fig. 4c).

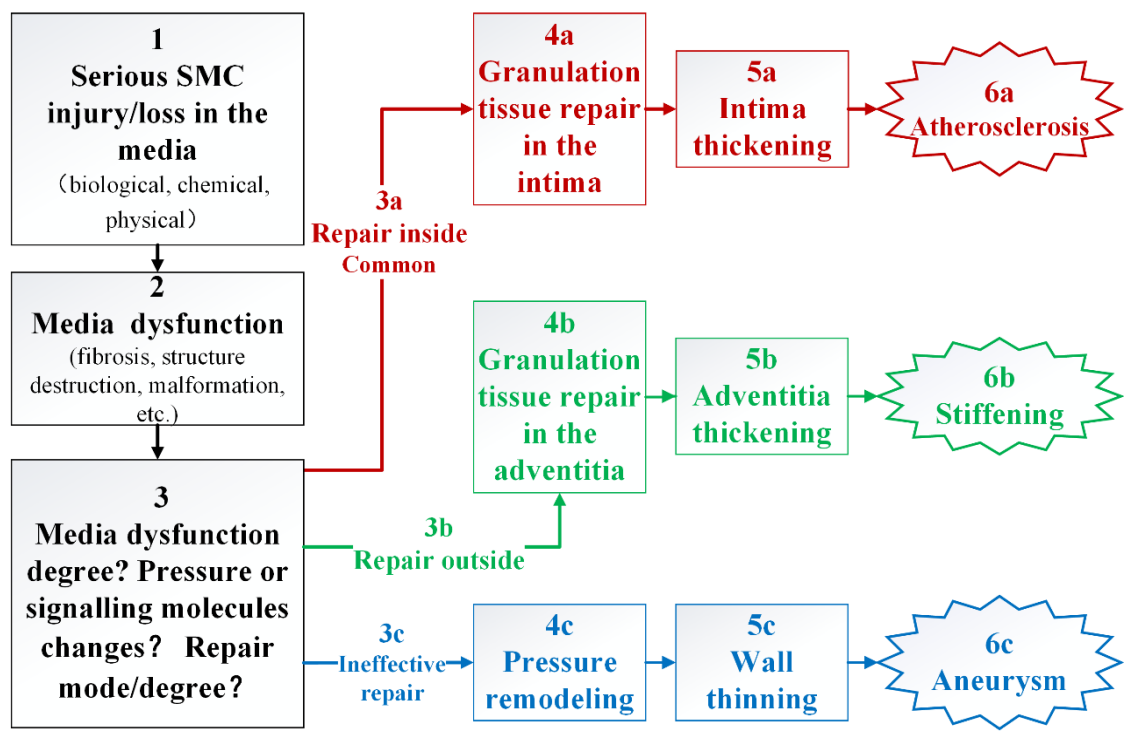

Fig.3 | Illustration of the theory

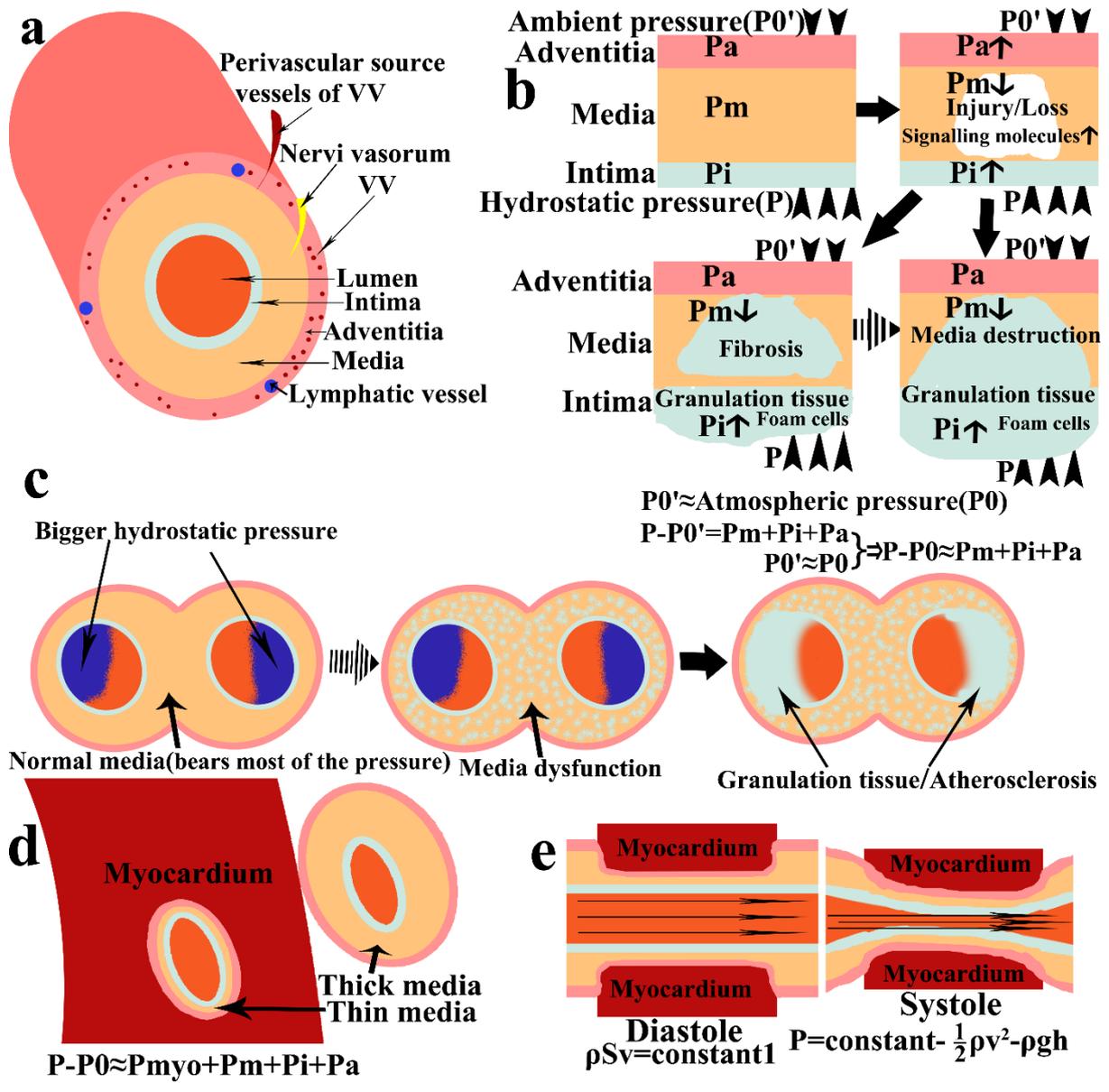

Fig.4 |

a: Structural diagram of normal muscular artery b: In clinical practice, measurement of blood pressure is the difference between cuff pressure and atmospheric pressure (P0). Hydrostatic pressure(P)-Ambient pressure(P0') = Pressure on vessel (media, intima, adventitia, etc.). In an individual, the ambient pressure(P0') ≈atmospheric pressure(P0). Therefore, hydrostatic pressure(P)- atmospheric pressure(P0) ≈Pressure on blood vessels (media, intima, adventitia, etc.). Media dysfunction leads to the mechanical property changes on the vessel, mainly the pressure redistribution. Serious media dysfunction leads to much more pressure increase on single cells of intima, media and adventitia for a significant reduction in the total number of cells. Media dysfunction causes the pressure(P-P0) undertaken by the media ($P_m$) reduced a lot, and the pressure undertaken by the intima ($P_i$) and adventitia ($P_a$) increased a lot which results in the myofibroblasts proliferation and secretion in the intima/adventitia (intima thickening is illustrated here). These lead to the granulation tissue repair in the intima and atherosclerosis. Granulation tissue repair in the intima is common for the richer nutrients in the lumen. Granulation tissue repair could sometimes occur in the adventitia where nutrients are enough, which was not illustrated here.

c: It is the illustration of bifurcations, curved arteries or lower extremity arteries where and why atherosclerosis is prone to occur. Hydrostatic pressure (P) is usually much bigger in the outer walls of bifurcations, the convex surface of curved arteries(2) or lower extremity arteries with Bernoulli's equation. When the media function is normal, it would not lead to much more pressure on the intima and adventitia in these "Relatively Vulnerable Zones" (RVZ) for the bigger ratio of SMCs to total cells or relatively much thicker media which shares the most part of pressure (P-P0) (Large denominator). However, if media function is seriously compromised in these arteries, the bigger pressure on the intima and adventitia in RVZ would be magnified (Denominator becomes much smaller). This leads to more severe lesions in RVZ.

d: It illustrated the hydrostatic pressure (P) and structural characteristics of myocardial bridge. The surrounding myocardium ($P_{myo}$) shares lots of pressure (P-P0), which leads to pressure reduced a lot on intima ($P_i$), media ($P_m$) and adventitia ($P_a$). This results in the very thin media of myocardial bridges. Even though the media was damaged, it would not lead to much more pressure on single cell of intima, media and adventitia for the much lower ratio of SMCs from total cells, which results in rare of atherosclerosis.

e: Effects of heart rhythm on hydrostatic pressure (P) of myocardial bridge. In systolic phase, blood velocity is bigger in myocardial bridge with the decrease of cross section according to ρSv=constant1 compared with that of the non-myocardial bridge. Hydrostatic pressure (P) is smaller in myocardial bridge with the increase of blood velocity according to P=constant-$\frac{1}{2}$ρv$^2$-ρgh compared with that of non-myocardial bridge during systole. Pressure on intima ($P_i$), media ($P_m$) and adventitia ($P_a$) of myocardial bridges further decreases than that of the non-myocardial bridges during systole.

## Summary paragraph

Atherosclerosis is the most common cause of cardiovascular and cerebrovascular diseases, although its pathogenesis remains unclear. Classical hypotheses, e.g., intimal injury, lipid deposition, low shear stress and turbulence, have been mainly geared towards atherogenesis. Unfortunately, a given hypothesis may only explain one or several phenomena. Our previous cognition failed to grasp its nature. Here we show that media dysfunction causes destruction in mechanical properties of media and leads to much more pressure redistributed on the intima and adventitia. Granulation tissue in the intima/adventitia could form to repair the media. Atherosclerosis, stiffening or aneurysm may develop pending upon the degree or mode of media dysfunction and granulation tissue repair of the onset. Nearly all characteristics of clinical atherosclerosis may be explained based on media dysfunction and bigger hydrostatic pressure (P) with our theory. We believe that media dysfunction is a key initiator and hemodynamic bigger hydrostatic pressure (P) aggravates atherogenesis(2). Measures, which could reduce SMCs injury/loss, preserve structure and function of the media or precise regulation of granulation tissue repair, should serve as potential candidates for preventing and curing atherosclerosis from etiology in the follow-up studies.

**Discussion**

Hemodynamic bigger hydrostatic pressure aggravates atherosclerosis(2). Except for capillaries and venules, all vessels possess three layers: intima, media and adventitia(3). As a continuous pipeline in human, the main differences in vessels lie in the medial layer bearing mechanical forces(3), especially the hydrostatic pressure (P) in muscular arteries possess relatively thicker media. Structural differences of vessels thus play a crucial role in human atherogenesis, while lipids, sugar, oxygen, and inflammatory factors, etc. are the participants (Incidence and degree of atherogenesis in human: Muscular arteries > Large elastic arteries > Arterioles; SVGs > RAGs). There are different resident cells in different layers(3). Intima comprises of endothelial cells and fibroblasts, etc. Media, on the other hand, mainly contains SMCs and few fibroblasts. Adventitia has fibroblasts, etc. According to the regeneration ability, they are roughly classified into labile cells or stable cells. Endothelial cells and fibroblasts, etc. belong to labile cells with strong proliferation, migration and repair abilities. SMCs are stable cells with weak capacity of proliferation and migration(4). Furthermore, in normal physiological conditions, nutrients for SMCs mainly come from diffusion of blood in the lumen and vasa vasorum (VV)(3). Therefore, in the presence of common cardiovascular risk factors including microcirculatory disturbance, inflammation, aging, smoking, mechanical injury or deranged metabolism, SMCs injury/loss(5) is more readily to develop which induces structural and functional damages of the media compared with the labile cells. In general, media, which is much more vulnerable than the intima and adventitia, are more prone to common pathological insults before fibrosis or structure destruction is developed, leading to media dysfunction. Media dysfunction causes destruction in mechanical properties of blood vessels and results in elevated pressure on single cell of intima, media and adventitia for a significant reduction in total cell abundance. With the vast pressure rise or signaling molecules release after media injury, myofibroblasts could be transformed from other cells, such as endothelial cells, fibroblasts, SMCs, etc. Secondary pathophysiological lesions would occur. According to the results and above evidences, the following theory was proposed.

**Media dysfunction is a key initiator in atherosclerosis**

Factors, e.g., biological, chemical, physical, induce serious injury/loss of SMCs in the media（Step 1, Fig.1, Extended Data Fig. 1）. Serious injury/loss of SMCs(5) leads to fibrosis (Fig.2c, Extended Data Fig. 4b) or structure destruction in the media (Fig.1b, Fig. 2a, 2c, Extended Data Fig. 2a, 3a, 4c), which results in media dysfunction（Step 2）. Media dysfunction leads to much bigger pressure on single cell of intima, media and adventitia. SMCs injury also leads to signalling molecules release. Much more pressure or signalling molecules activate endothelial cells, fibroblasts, SMCs, etc. into myofibroblasts(6, 7). Myofibroblasts secrete abundant ECM which forms extracellular collagen fibers(6, 8)(Fig.1, Fig.2a; Extended Data Fig. 1b, 2, 3), foam cells and extracellular lipid accumulation[7]. In vascular tissues, myofibroblasts, ECM and lumen (intima granulation tissue, Fig.1a)/VV (adventitia granulation tissue, Fig.2a) constitute the granulation tissue repair. Media dysfunction degree and the repair mode/degree determine secondary lesions, e.g., atherosclerosis (Step 3a), stiffening (Step 3b) or aneurysm (Step 3c). Proliferation and secretion of myofibroblasts in the intima (Step 4a, repair inside)/adventitia (Step 4b, repair outside) leads to intima thickening (Step 5a,

common)/adventitia thickening (Step 5b). Intima thickening (Fig.1), a repair inside pattern, is a common repair for the relatively richer nutrients in the lumen compared with the adventitia. While adventitia thickening (Fig.2a) sometimes occurs in the vessels whose microcirculation is well in the adventitia (Extended Data Fig.3a). As the intima thickening, the lumen would be gradually unable to provide enough nutrients, which could induce angiogenesis(9) (Fig.1b, Extended Data Fig.2a). With granulation tissue hyperplasia, lots of collagen fibers and foam cells form in the intima[7] (Fig.1b, Fig.2a, Extended Data Fig.2, Extended Data Fig.3). There are also lots of collagen fibers are denaturing gradually[7] (Extended Data Fig.2). As the intima thickening, it would aggravate ischemia and hypoxia. This forms a vicious circle and eventually results in atherosclerosis (Step 6a). Repair outside would not narrow the lumen and might be one reason for vessel stiffening (Step 6b). If such repair is ineffective (Step 3c), aneurysm (Step 6c) may be gradually developed with the arterial pressure. Atherosclerosis, stiffening or aneurysm develops depending on the degree or mode of media dysfunction and granulation tissue repair upon onset of pathology. There are many differences between SVGs and arteries in atherogenesis for the very thin media of SVGs (Fig.3, Fig.4a, Fig.4b).

**Extensive and serious media dysfunction in SVGs**

In patients underwent CABG, SVGs are prone to aneurysm or extensive and serious atherosclerosis. The patency of RAGs is much better than that of SVGs(1). These phenomena are unable to be interpreted using intima injury, lipid deposition, low shear stress, turbulence or inflammation. Atherosclerosis is unique between SVGs and RAGs for differences in media dysfunction. SVGs with very thin media (extensive and serious media dysfunction), aneurysm or atherosclerosis could form easily depending on granulation tissue repair effect. Although RAGs lost VV microcirculation which might induce some injury in the adventitia and media, blood in the lumen could still nourish most of the thick media, which could still undertake most pressure (P-P0), and would not lead to bigger pressure increment and serious granulation tissue repair in the intima (Fig.3). The patency of RAGs is much better than that of SVGs in CABG patients(1), and the results also demonstrated the protective effect of thicker media of RAGs and the aggravating effect of the thinner media of SVGs (Media dysfunction) in atherogenesis.

**Variation in susceptibility in different blood vessels**

In human, the medium and large arteries, especially the muscular arteries, are prone to atherogenesis, while arterioles and in situ veins are usually spared. The media is relatively thicker, the bigger the ratio of SMCs to total cells (Fig.4a; Fig4b). When the hydrostatic pressure (P) is similar, the more SMCs were damaged, pressure increased much more on single cell of intima, media and adventitia (Fig.4b). The relatively thicker media serve as both pressure bearer (protective effect) and vulnerable targets for pathological factors. The higher the degree of damage, the more serious media dysfunction it is and the more pressure must be transferred to the intima and adventitia (Fig.4b). Large elastic arteries usually have well developed adventitia with thick and tough connective tissue which undertakes lots of pressure(P-P0) from the lumen and is less vulnerable. Therefore, the media is relatively thin compared with muscular arteries. Even though media were damaged, the thick and tough adventitia could still withhold pressure which is shared much pressure with intima and media, resulting in less increment of pressure in the intimal layer. As a result, fewer myofibroblasts are formed, leading to mild intima thickening and atherosclerosis compared with the muscular arteries with relatively thick media and thin adventitia. Since pressure (P-P0) is shared or low in them, arterioles and in situ veins have little pressure in the media, intima and adventitia, resulting in thin media. Even though the media was damaged, it would not lead to overt pressure elevation on single cell of intima, media and adventitia, not favoring onset of atherosclerosis.

**Hydrostatic pressure is usually bigger in the outer walls of bifurcations, the convex surface of curved arteries or lower extremity arteries**

Atherosclerosis is usually more pronounced in the outer walls of bifurcations, convex surface of curved arteries or lower extremity arteries. Hemodynamic bigger hydrostatic pressure aggravates atherosclerosis(2). Hydrostatic pressure (P) is usually much bigger in the outer walls of bifurcations, the convex surface of curved arteries(2) or lower extremity arteries (Fig.4c). When the arteries do not suffer from media dysfunction, the hydrostatic pressure(P) increase in "Relatively Vulnerable Zones" (RVZ) would not lead to atherosclerosis. For the normal media bearing most of the pressure(P-P0) for the relatively thick media which has a much bigger ratio of SMCs to total cells (Denominator is very large), the pressure increase on the intima/adventitia is not obvious in RVZ without media dysfunction. However, if serious SMCs injury/loss is in the media, the pressure increase on the single cell of RVZ would be much bigger than that of their counterparts for the total cells decreased significantly (Denominator decreases significantly). With our theory, much bigger hydrostatic pressure could aggravate lesions, e.g., atherosclerosis, stiffening or aneurysm (Fig.3; Fig.4b). Therefore, bigger hydrostatic pressure (P) in RVZ is an important aggravating factor rather than a determinant. Normal media function is the decisive protection factor and media dysfunction is the determinant of the lesions in the muscular arteries with relatively thicker media, thin intima and loose adventitia.

**Myocardial bridges are rare with atherosclerosis**

Myocardial bridges are existed in approximately 1/3 of adults[2]. Interestingly, the myocardial bridges are much less likely to undergo atherosclerosis compared with the non-myocardial bridges in an individual(10, 11). The surrounding myocardium ($P_{myo}$) shares much pressure (P-P0), which leads to much lower pressure on intima, media and adventitia compared with their counterparts (Fig.4d). Most of myocardial bridges do not suffer from obvious myocardial ischemia[2]. This means that they have little effect on blood perfusion in normal activity. Therefore, fluid continuity equation could be used in them. Because of the myocardium around the artery, sectional area of myocardial bridge becomes smaller in the systole. With fluid continuity equation: ρSv=constant1 (ρ= blood density, S= sectional area, v= blood velocity). Blood can be considered incompressible and ρ is a constant in an individual. Blood velocity is inversely proportional to the sectional area. Blood velocity is bigger in myocardial bridges with sectional area decrease compared with non-myocardial bridges in the systole. According to P=constant-$\frac{1}{2}$ρv$^2$-ρgh, the bigger the blood velocity, the smaller the hydrostatic pressure(P). This further leads to smaller pressure on the intima, media and adventitia in the myocardial bridges as opposed to the non-myocardial bridges during systole (Fig.4e). These reasons cause the very thin media of myocardial bridges. Even the thin media was completely damaged by the common pathological factors, it would not lead to much more pressure on single cell of intima, media and adventitia, which results in rare of atherosclerosis. (Fig.4d; Fig.4e).

**Relative media dysfunction in severe hypertension**

The incidence rate of coronary heart disease is closely related to that of hypertension.

Reminiscent of a balloon full of air, artery without media dysfunction could adapt to appropriate pressure(P-P0) changes. Not a lot of pressure increase would not lead to much more pressure on the intima/adventitia for the thick media sharing the most pressure (P-P0). However, if pressure (P-P0) is too big for artery to bear, the media, intima and adventitia must undertake much more pressure. In a sense, severe hypertension leads to relative media dysfunction. On the one hand, proliferation of SMCs is too weak(4) for the media to compensate; on the other, pressure increasing a lot in the intima/adventitia could induce myofibroblasts formation, proliferation and secretion, which lead to granulation tissue repair and intima/adventitia thickening. Secondary lesions would form with our theory.

**Conclusions**

Almost all characteristics of clinical atherosclerosis could be interpreted with our theory. We believe that media dysfunction is a key initiator in the pathogenesis of atherosclerosis. Hemodynamic bigger hydrostatic pressure aggravates the atherosclerotic pathological process(2). Furthermore, atherosclerosis, stiffening or aneurysm should share the same initiating factor namely media dysfunction. Measures capable of reducing SMCs injury/loss, preserving structure and function of the media or the precise regulation of granulation tissue repair, thus offering promises in the prevention and treatment of atherosclerosis. This theory should help to address the perplexing issues around over the past 150 years. We believe that our theory should provide new avenue in the treatment of atherosclerosis from etiology in the future.

**Acknowledgements**

This work was supported by the China Postdoctoral Science Foundation (2018M641934), a grant of Innovative Research Groups of the National Natural Science Foundation of China (81521001), a grant to A. Sun from the National Science Fund for Distinguished Young Scholars (81725002), the National Natural Science Foundation of China (81570224), and the Major Research Plan of the National Natural Science Foundation of China (91639104). Thanks for Zhengdong Li from Institute of Forensic Sciences of Shanghai providing the autopsy specimens.

**Author contributions**

Junbo Ge supervised the study. Xinggang Wang and Junbo Ge conceived and designed the experiments. Xinggang Wang performed experiments, analyzed data, and wrote the manuscript. Junbo Ge, Aijun Sun made manuscript revisions.

**Competing interests**

The authors declared that they have no conflicts of interest to this work.

# Supplementary materials

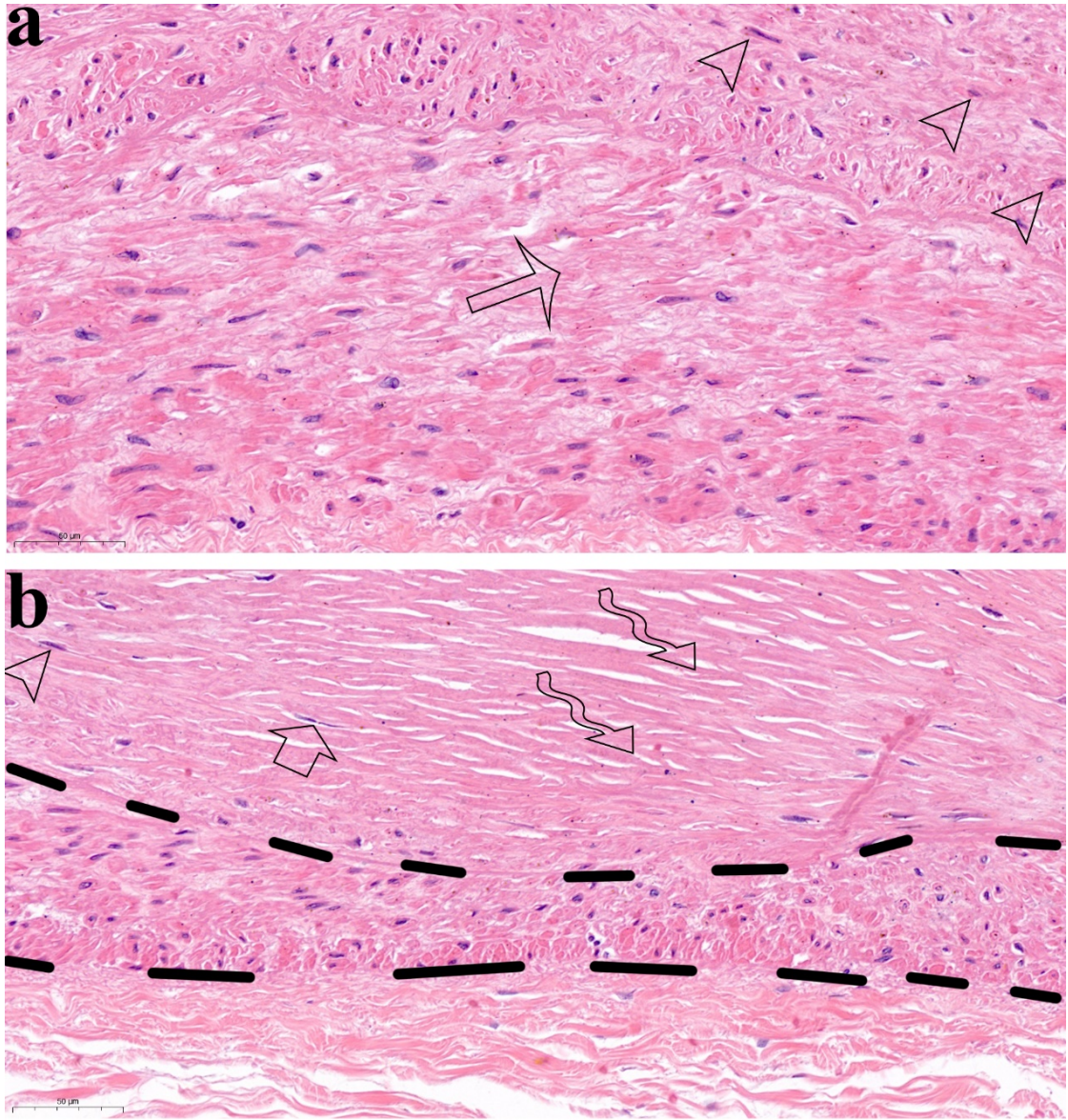

Extended Data Fig. 1 | Media dysfunction and granulation tissue repair inside in the early stage (fibrous plaque)

It is the high-power fields of Fig.1a. The triangle of media, at the beginning of media dysfunction, was a true portraiture of cells injury, while the triangle of intima was a true portraiture of myofibroblasts formation, secretion of ECM and collagen fibers accumulation in the intima along with the media dysfunction. In the triangle of media, the media became thicker with cells injury(a). Cells structure loss and nucleus disappearance are the signs of cells injury (arrow). In the intima closing to the media with cells injury, myofibroblasts with long spindle nucleus and eosin deeply stained cytoplasm (arrow head) were secreting ECM. Myofibroblasts proliferation and ECM secretion need lots of nutrients. In non-vascular tissues, myofibroblasts/fibroblasts, ECM and capillaries constitute granulation tissue repair. In vascular tissues, nutrients diffusion from lumen blood replaced the functions of capillaries and could provide enough nutrients for myofibroblasts proliferation and ECM secretion from the beginning. Therefore, in vascular tissues, myofibroblasts, ECM and lumen (intima)/capillaries(adventitia) constitute the granulation tissue repair from the beginning. With the repair development, ECM formed lots of extracellular collagen fibers (b, curved arrow) with shuttle like staggered arrangement. The media became much thinner and was repaired by the granulation tissue (b).

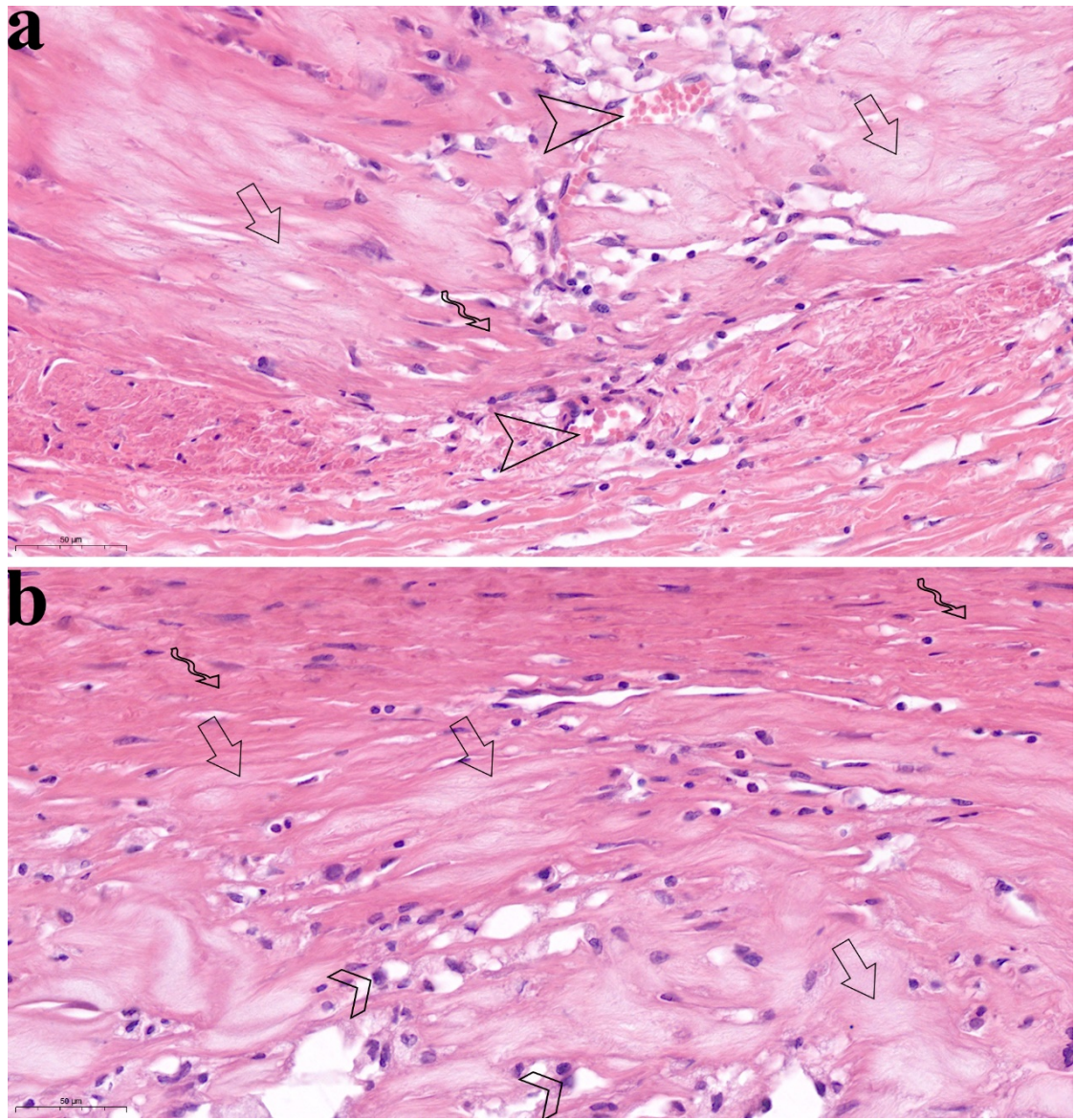

Extended Data Fig. 2 | Media dysfunction and granulation tissue repair inside in the later stage (soft plaque)

It is the high-power fields of Fig.1b. HE shows that thick granulation tissue repaired the serious destroyed media which lost structure and function(a). There was angiogenesis (arrow head) in the granulation tissue. New vessels passed through the serious injured media to supply small parts of the thick granulation tissue near the injured media. And the normal collagen fibers with eosin stained (a, b, curved arrow) were distributed in the areas with new capillaries or near the lumen. Eosin light stained (a, b, straight arrow) were denatured collagen fibers, and some of them closing to the normal collagen fibers still preserved the shape of collagen fibers. It means that they were denaturing gradually (a, b). Lots of denatured collagen fibers were widely distributed in the areas without new capillaries and far from the lumen (a, b). There were also lots of foam cells around the fibers in the granulation tissue(b).

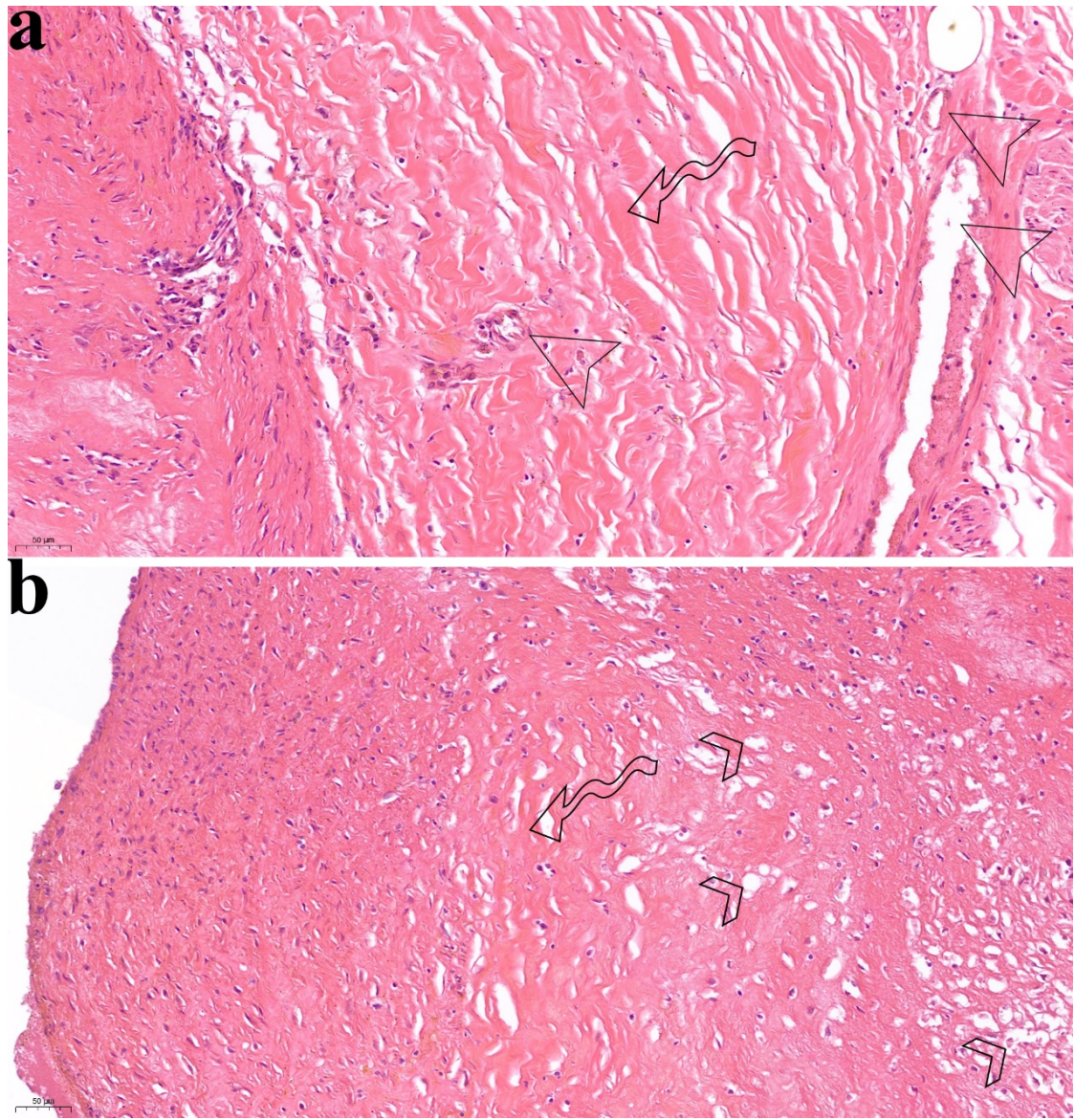

Extended Data Fig. 3 | Media dysfunction and granulation tissue repair in both intima and adventitia

It is the high-power fields of Fig.2a. HE shows that the media was destroyed seriously and the lumen had severe stenosis. Granulation tissue repair could be seen in both adventitia(a) and intima(b). Lots of collagen fibers were widely distributed in both adventitia and intima (curved arrow). There were many small vessels nourishing the granulation tissue around the adventitia (a, arrow head). Foam cells were widely distributed in the intima (b, arrow head).

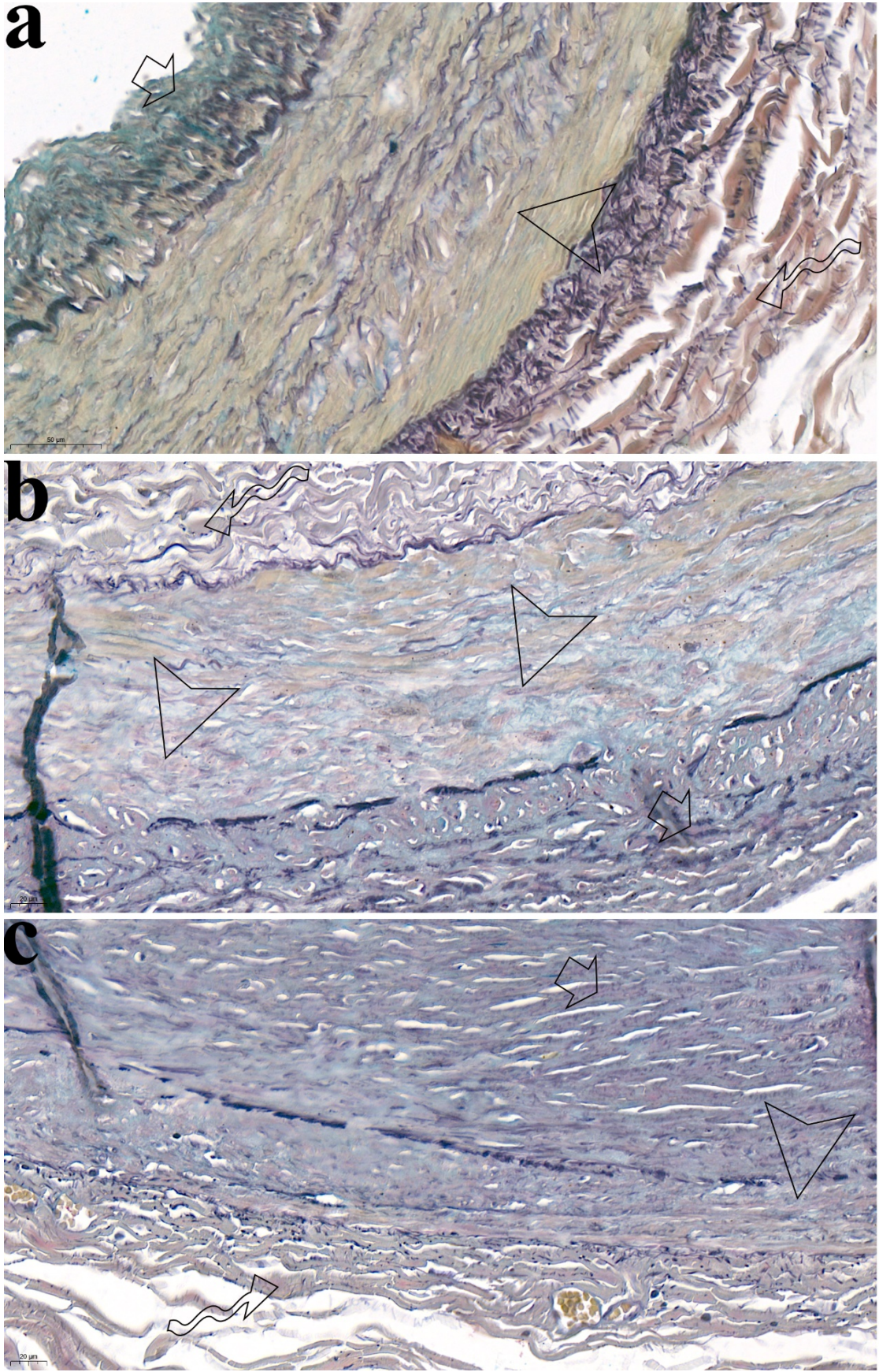

Extended Data Fig. 4 | Muscular property loss, fibrosis or structural damage of media in artery

with atherosclerosis

a: It is the high-power fields of Fig.2b. MOVAT staining showed that the media of normal coronary artery had good muscular property: The media (arrow head) was the main part of the wall and was thick and dense compared with the intima (straight arrow) and adventitia (curved arrow). The intima was thin and the adventitia was loose. The media was constituted of mainly muscle tissues stained with yellow, few elastic fibers stained with black and few proteoglycans stained with bluish green.

b: It is the high-power fields of Fig.2c. MOVAT staining showed that coronary artery, which suffered from muscular property loss and fibrosis in the media, had fibrous plaque in the intima. The media lost muscular property with few muscle tissues stained with yellow, many proteoglycans stained with bluish green and many collagen fibers stained with red (arrow head). There was mild atherosclerosis with collagen fibers in the intima (straight arrow). There were lots of loose collagen fibers in the adventitia (curved arrow).

c: It is the high-power fields of Fig.2c (serious media dysfunction). MOVAT staining showed that the media of coronary artery suffered from serious muscular property loss, thinning and structural damage. The media was thin, had almost no yellow stained muscle tissue and lost function (arrow head). In the intima, atherosclerosis was serious (straight arrow).